\documentclass[twocolumn,twocolappendix]{aastex701}
\usepackage{amsmath}
\usepackage{lineno}

\graphicspath{{./}{figures/}}

\submitjournal{ApJ}

\shorttitle{Stellar Evolution in AGN Disks}
\shortauthors{Dittmann \& Cantiello}

\begin{document}
\title{The Effects of Accretion Feedback on Stellar Evolution in AGN Disks}

\correspondingauthor{Alexander J. Dittmann}
\email{dittmann@ias.edu}

\author[0000-0001-6157-6722]{Alexander J. Dittmann}
\affiliation{Institute for Advanced Study, 1 Einstein Drive, Princeton, NJ 08540, USA}
\altaffiliation{NASA Einstein Fellow}
\email{dittmann@ias.edu}

\author[0000-0002-8171-8596]{Matteo Cantiello}
\affil{Center for Computational Astrophysics, Flatiron Institute, 162 5th Avenue, New York, NY 10010, USA}
\affil{Department of Astrophysical Sciences, Princeton University, Princeton, NJ 08544, USA}
\email{mcantiello@flatironinstitute.org}

\begin{abstract}
Stars embedded in the accretion disks of active galactic nuclei (AGN) can accrete rapidly from their surroundings, dramatically altering their structure and evolution. However, feedback from the release of gravitational potential energy and radiative enthalpy by accreting gas can limit accretion rates, as recently demonstrated in radiation hydrodynamics simulations. To determine the importance of these effects neglected in earlier stellar evolution calculations, we incorporate these feedback processes into a semi-analytical model of stellar structure and evolution and conduct a suite of calculations spanning a broad parameter space of AGN disk conditions drawn from $\alpha$-disk models with central black hole masses $M_\bullet/M_\odot \in [10^6, 10^9]$. We find that accretion feedback limits stellar accretion rates below $\sim 10^{-1}\,M_\odot\,\mathrm{yr}^{-1}$, reducing the sensitivity of stellar evolution on disk properties. This suppression eliminates runaway accretion in models where it would otherwise occur, broadening the parameter space over which stars can reach long-lived ``immortal'' equilibria between accretion and mass loss. When gap opening is also accounted for, accretion feedback significantly alters stellar properties: it can reduce accretion and mass-loss rates by over an order of magnitude, reducing the strength of accretion shocks and thereby increasing equilibrium stellar masses and radii. These higher masses correspond to higher intrinsic luminosities, suggesting that neglecting accretion feedback may lead to an underestimate of disk chemical enrichment rates. Additionally, accretion feedback is important for predicting the properties of stellar populations within AGN disks, and associated transient phenomena. 
\end{abstract}


\section{Introduction} \label{sec:intro}

Active galactic nuclei (AGN) are fed by vast reservoirs of gas, which accrete onto a central supermassive black hole (SMBH) and in the process release enormous energy \citep[e.g.,][]{1969Natur.223..690L,2008ARA&A..46..475H}. The outer regions of these accretion disks may be gravitationally unstable, globally or in local patches, leading to the formation of stars and compact objects \citep[e.g.,][]{1980SvAL....6..357K,2003MNRAS.339..937G,2023MNRAS.521.4522D,2024OJAp....7E..71H}. If a nuclear star cluster is present before an accretion episode begins, those stars may be dragged into the disk by a number of processes such as dynamical friction and the excitation of bending waves throughout the disk \citep[e.g.,][]{1993ApJ...409..592A,2020MNRAS.499.2608F}. Regardless of their origin, these stars can be dramatically altered by the dense environments surrounding them.

Stellar populations within AGN disks are thought to play an crucial role in enriching their disks with fusion byproducts \citep[e.g.,][]{1993ApJ...409..592A,2022ApJ...929..133J}, inspired by the apparently high metal content of AGN relative to quiescent galaxies \citep[e.g.,][]{1999ARA&A..37..487H,2018MNRAS.480..345X}. Our own Galactic center might even retain a few dynamically and chemically peculiar disk-affected stars, left over from the most recent accretion episode onto Sagittarius A* \citep[e.g.,][]{Levin:2003,Do:2009,Habibi:2017}.
Stars in and around AGN disks are also thought to cause extreme flares, many of which are thought to result from tidal disruptions or superluminous supernovae as the gaseous disk may help convert kinetic energy from those events into radiation \citep[e.g.,][]{2017MNRAS.470.4112G,2021ApJ...920...56F}. Additionally, stellar remnants embedded within the AGN disks may be substantial sources of gravitational wave events, both for stellar-mass mergers \citep[e.g.,][]{2017MNRAS.464..946S,2020MNRAS.498.4088M} and extreme-mass-ratio inspirals \citep[e.g.,][]{2021PhRvD.103j3018P,2023MNRAS.521.4522D}.

Pioneering studies of stellar evolution in AGN disks \citep[e.g.,][]{2021ApJ...910...94C} made numerous simplifying assumptions in addition to the limitations intrinsic to one-dimensional stellar evolution calculations. Major uncertainties include how stars within AGN disks both accrete and loose mass, how accretion is altered when the pressure gradients resulting from a star's radiation field becomes comparable to its gravitational pull, and how the accretion stream itself affects the structure and evolution of the star.
Subsequent stellar evolution studies investigated geometric and tidal effects \citep{2021ApJ...916...48D}, the effects of stellar rotation and the accretion of angular momentum \citep{2021ApJ...914..105J}, the importance of the composition of the accreting material \citep{2023ApJ...946...56D}, and the significance of various assumptions concerning the efficiency of compositional mixing in the stellar interior \citep{2023MNRAS.526.5824A}.

Following those earlier studies, \citet{2024ApJ...974..106C,2025ApJ...987..188C} made progress in understanding the interplay between accretion streams and strong radiation fields by performing a small suite of multidimensional radiation hydrodynamics simulations of stars accreting from dense disks of gas. These simulations demonstrated that in some regimes, the accretion rate onto the star can be limited by feedback from the release of radiative enthalpy and potential energy from the accreting gas. Some preliminary stellar structure and evolution calculations in \citet{2026ApJ...997..206X} took inspiration from these results, but were unable to implement these limits on accretion directly or survey a wide parameter space.\footnote{Specifically, Equations 19 and 20 of \citet{2026ApJ...997..206X} scale the accretion rate by a factor of $[1 - (L_*/L_0)^8/(1 + (L_*/L_0)^8)]^2$, where $L_*$ is the stellar luminosity and $L_0$ is a fraction (0.25 to 0.9) of the Eddington luminosity, shutting off accretion artificially and leaving it unclear whether feedback processes or the aforementioned ad hoc limit played a significant role in stellar evolution.}

In this work, we investigate the effects of these feedback processes using the semi-analytical model of stellar structure and evolution presented in \citet{2025ApJ...979..245D}, which combines the analytical models of stellar structure developed in \citet{1926ics..book.....E} and \citet{1984ApJ...280..825B} with various accretion and mass-loss models developed for modeling stars in AGN disks. This semi-analytical model assumes the compositional uniformity of the star and a polytropic structure, and is therefore inaccurate for low-mass stars but well-suited to the high-mass nearly-fully-convective stars typical in AGN disks; examples shown in \citet{2025ApJ...979..245D}, particularly Figures 2 and 5, highlight the agreement of this simple semi-analytical model with suites of full stellar structure and evolution calculations using \texttt{MESA} \citep{2011ApJS..192....3P}. In this work, we extend this model to include the limitation of accretion by radiative enthalpy and potential energy feedback, largely following \citet{2024ApJ...974..106C}. We also conduct a large suite of calculations across a suite of AGN disk models (following \citet{2003MNRAS.341..501S}), to examine the role of these various models for accretion across a range of AGN-like conditions. 

In Section \ref{sec:methods} we summarize our basic model and the various assumptions we make about mass loss and accretion onto stars in AGN disks. In Section \ref{sec:results} we present a few pedagogical calculations, followed by a broad survey over a wide range of AGN-disk-like conditions. We discuss some implications and limitations of our results in Section \ref{sec:discussion}, and conclude in Section \ref{sec:conclusion}. Appendix \ref{app:escape} describes ancillary modifications we have made to our semi-analytical model of stellar structure since we presented it in \citet{2025ApJ...979..245D}, and Appendix \ref{app:diskModels} describes in some detail the AGN disk models used in this work. 

An implementation of our model is available at \href{https://github.com/ajdittmann/starsam/tree/dev}{https://github.com/ajdittmann/starsam/tree/dev}.
\section{Methods}\label{sec:methods}
Our approximation of stellar structure, following \citet{1926ics..book.....E,1984ApJ...280..825B}, is the same as described in Section 2.1 of \citet{2025ApJ...979..245D};  crucially, this model provides values for the stellar radius ($R_*$) and intrinsic luminosity ($L_*$, resulting from nuclear fusion) as functions of the stellar mass ($M_*$) and composition (mass fractions $X_*$, $Y_*$, and $Z_*$). The evolutionary variables are simply the mass of hydrogen, helium, and metals constituting the star, which change due to mass loss, accretion, and nuclear fusion. Section \ref{sec:basicModel} describes our models for stellar accretion under various approximations. Section \ref{sec:massLoss} briefly describes our approximation for stellar winds. See Section 2 of \citet{2025ApJ...979..245D} for additional details.

\subsection{Models for Stellar Accretion}\label{sec:basicModel}
The essential ingredients required to model the evolution of stars within AGN disks are the stellar accretion rate $\dot{M}_+$ and mass-loss rate $\dot{M}_-$. 
The accretion rate onto the star is determined by the conditions of the accretion disk outside the star, the tidal influence of the central supermassive black hole, and how strongly the star perturbs the AGN disk around it, in addition to the radiative feedback of the star on its surroundings and other sources of feedback on the accretion stream. To make notation more concrete, we begin by considering a variety of base accretion rates ($\dot{M}_0$), not considering any feedback; we use this as an input to calculate a stellar-radiation-limited accretion rate $\dot{M}_L$, and an accretion-feedback-limited accretion rate $\dot{M}_f$, which determine the ultimate rate of accretion onto the star ($\dot{M}_+$).

Dimensionally, accretion rates can be expressed as the product of the area of a surface, a characteristic density, and the characteristic velocity of the matter passing through that surface. In the simplest case, applicable to the lowest-mass stars, the relevant speed is that of sound waves in the ambient medium, the characteristic density that of the AGN disk in the vicinity of the star, and the relevant surface is that of a sphere with radius $R_B=2GM_*/c_s^2$, leading to an accretion rate of
\begin{equation}\label{eq:bondi}
\dot{M}_B \approx \pi \rho c_s R_B^2,
\end{equation}\
which we refer to as the Bondi rate \citep{1952MNRAS.112..195B}. When considering no other effects, we by default take $\dot{M}_0=\dot{M}_B$, the ``Bondi'' rate.

Since $R_B\propto M_*^2$, as stellar masses grow this length scale can quickly become comparable to or exceed that over which the gravitational pull of the star can overcome the tidal gravitational field of the central supermassive black hole. The latter length scale is the Hill radius \citep{hill1878researches}, given by 
\begin{equation}
R_H = r\left(\frac{M_*}{3M_\bullet}\right)^{1/3}=\left(\frac{GM_*}{\Omega^2}\right)^{1/3},
\end{equation}
where $r$ is the distance from the star to the SMBH, $M_\bullet$ is the mass of the central SMBH, and $\Omega$ is the angular velocity of the stellar orbit through the AGN disk. As stars grow, this scale can limit the supply of matter available for accretion, leading to an accretion rate \citep[e.g.,][]{2020MNRAS.498.2054R}
\begin{equation}
\dot{M}_H\approx \pi \rho c_s R_H^2.
\end{equation}
Accounting for tidal effects then, a general expression for the base accretion rate is $\dot{M}_0=\pi \rho c_s\min{(R_H^2,R_B^2)}$, the ``tidal'' rate.

If stars become sufficiently massive, their gravitational influence can begin to reshape the AGN disk surrounding them, carving a deep gap and reducing the average density of gas near the star. The surface density near the star, $\Sigma'$, relative to its unperturbed value $\Sigma_0$, is given by \citep[e.g.,][]{2017PASJ...69...97K,2020ApJ...891..108D}
\begin{equation}
\xi\equiv\frac{\Sigma}{\Sigma_0} = \frac{1}{1+0.04K},
\end{equation}
where
\begin{equation}
K\equiv \alpha^{-1}\left(\frac{M_*}{M_\bullet}\right)^2\left(\frac{H}{r}\right)^{-5}
\end{equation}
quantifies the ratio between the viscous angular momentum flux near the star and the torque on the star by the outer disk \citep[e.g.,][]{1986ApJ...309..846L,2014ApJ...782...88F,2017PASJ...69...97K,2020ApJ...891..108D}. The accretion rate in this gap-opening regime can be approximated simply as 
$\dot{M}_0=~\xi\pi\rho c_s \min{(R_H^2,R_B^2)}$ \citep[e.g.][]{2023MNRAS.525.2806C}, the ``gap'' accretion rate.

The aforementioned accretion rates are fairly accurate in the absence of feedback processes. However, once rate of momentum transfer to gas by the impingement of stellar radiation upon it becomes comparable to the gravitational pull of the star, the accretion process should be severely modified. We quantify this by comparing the stellar luminosity to the Eddington luminosity, which we define as 
\begin{equation}
L_{\rm Edd}=6\times10^{37}(1+X)(M_*/M_\odot)\rm{erg~s^{-1}},
\end{equation}
assuming electron scattering to be the relevant source of opacity and using $X$ to represent the hydrogen mass fraction of the gas. In spherical symmetry, this radiation field would reduce the effective mass of the star, felt by the surrounding gas, by a factor of $(1-L/L_{\rm Edd})$; assuming Bondi accretion, the accretion rate would thus be reduced by a factor of $(1-L/L_{\rm Edd})^2$, or in the high-mass tidally limited case by $(1-L/L_{\rm Edd})^{2/3}$; these assumptions, among others, were explored in stellar structure and evolution calculations by \citet{2021ApJ...916...48D}. Alternatively, inspired by the latitudinal brightness variations of stellar surfaces \citep[e.g.,][]{1924MNRAS..84..665V,1967ZA.....65...89L} and the concurrence of inflow and outflow thanks to the Rayleigh-Taylor instability and other progenitors of turbulence \citep[e.g.][]{2013MNRAS.434.2329K,2014ApJ...796..107D,2019ApJ...885..144J}, \citet{2021ApJ...910...94C} adopted a smoother prescription that could allow some accretion even after the stellar luminosity exceeded the Eddington scale, 
\begin{equation}\label{eq:radFB}
\dot{M}_L = \dot{M}_0(1-\tanh|L/L_{\rm Edd}|).
\end{equation}
Equation (\ref{eq:radFB}) is the approximation we adopt in this work. 

We must clarify that in the preceding expressions, $L\equiv L_*+L_S$, the sum of the intrinsic stellar luminosity from nuclear fusion and a shock luminosity, which follows from the thermalization of the kinetic energy of the accretion stream \citep[e.g.,][Section 4.5]{2021ApJ...910...94C}. This leads to $L_S\equiv\dot{M}_+v_{\rm esc}^2$, where $v_{\rm esc}$ is the escape speed form the stellar surface. Previous works have take $v^2_{\rm esc}=2GM_*/R_*$. However, to avoid paradoxical results in the high-$\dot{M}_+$ low-$M_*$ regime explored in this paper, we reduce the effective mass of the star by a factor of $(1-L/L_{\rm Edd})$ when defining the escape velocity. We discuss this in more detail in Appendix \ref{app:escape}. 

\subsubsection{Additional Forms of Feedback}\label{sec:feedback}
Recently, \citet{2024ApJ...974..106C} demonstrated, by carrying out radiation hydrodynamics simulations of stellar accretion, that other types of feedback can reduce the rate of accretion onto a disk-embedded star even when the stellar luminosity is well below the Eddington scale. Moreover, \citet{2024ApJ...974..106C} found that simple analytical prescriptions matched the accretion rates measured in their radiation hydrodynamics simulations quite accurately.
Namely, \citet{2024ApJ...974..106C} argue that both the gravitational potential energy and the radiative enthalpy of accreting gas can contribute to a diffusive radiative luminosity within the accretion stream, leading to two limits on the accretion rate: one following from radiative enthalpy, which we implement as
\begin{equation}\label{eq:dotMr}
\dot{M}_r=\left(1-\frac{L}{L_{\rm Edd}} \right)\frac{L_{\rm Edd}}{4c_{s,r}^2},
\end{equation}
where $c_{s,r}^2\equiv P_{\rm rad}/\rho$ is the effective sound speed from radiation pressure and $4P_{\rm rad}/\rho$ is the specific radiative enthalpy,\footnote{Specifically, \citet{2024ApJ...974..106C} set $\dot{M}_r=(L_{\rm Edd}-L)\rho_0/(4aT_0^4)$, a factor of three smaller than the result in the optically thick limit. Intriguingly, Figure 6 of \citet{2024ApJ...974..106C} illustrates that in the radiative-enthalpy-limited regime, the formula used by \citet{2024ApJ...974..106C} for $\dot{M}_r$ underestimates the accretion rate measured from their simulations by a factor of $\sim3$.}
and another related to potential energy, which we implement as
\begin{equation}\label{eq:dotMg}
\dot{M}_g = \left(1-\frac{L}{L_{\rm Edd}} \right)\frac{2L_{\rm Edd}}{v_{\rm esc}^2}.
\end{equation}
Because we will investigate stellar evolution under these approximations over a wide range of conditions, we follow works such as \citet{2003MNRAS.341..501S} to define a radiation sound speed that quantifies the coupling between the fluid and the radiation field through the optical depth
\begin{equation}\label{eq:csr}
c_{s,r}^2 = \frac{\tau\sigma}{2c\rho}T^4\left(\frac{3}{8}\tau + \frac{1}{2} + \frac{1}{4\tau}\right)^{-1}.
\end{equation}

As suggested by \citet{2024ApJ...974..106C}, these estimates can be combined to define an accretion-feedback-limited accretion rate of 
\begin{equation}\label{eq:mdotFB}
\dot{M}_f = \dot{M}_0\!\left(\!1 - \frac{\dot{M}_f}{\dot{M}_{g}} - \frac{\dot{M}_f}{\dot{M}_r}\!\right)^2\!\!\approx\min{[\dot{M}_0,\dot{M}_g,\dot{M}_r]},
\end{equation}
where \citet{2024ApJ...974..106C} specifically assumed $\dot{M}_0=\dot{M}_B$ but we will consider more general accretion prescriptions. One feature of Equation (\ref{eq:mdotFB}) is that above certain limits, the accretion rate can become independent of the ambient density, and at higher rates independent of stellar mass.

When we consider these additional forms of feedback in our calculations, we take the accretion rate to be $\dot{M}_+ = \min{(\dot{M}_L,\dot{M}_f)},$ and otherwise simply set $\dot{M}_+=\dot{M}_L$.

\subsection{Mass Loss}\label{sec:massLoss}
In addition to accretion, the evolution of stars in AGN disks is governed by mass loss. Stellar mass loss, be it driven by the stellar continuum or various lines, especially in the near-Eddington regime relevant here, is a complex process, and nontrivial to model even in spherical symmetry \citep[e.g.,][]{2004ApJ...616..525O,2017MNRAS.472.3749O}, let alone in the presence of turbulence and in multiple dimensions \citep[e.g.][]{2018Natur.561..498J,2025arXiv250812486G}. We forego any attempt to model these processes precisely, and instead rely on a simple prescription qualitatively suitable for super-Eddington continuum-driven mass loss \citep[e.g.,][]{2011ApJS..192....3P,2021ApJ...910...94C},
\begin{equation}\label{eq:massloss}
\dot{M}_-=\frac{L}{v_{\rm esc}^2}\left[1 + \tanh{\left(\frac{L-L_{\rm Edd}}{0.1L_{\rm Edd}}\right)}\right].
\end{equation}

\section{Results}\label{sec:results}
Stellar evolution within AGN disks, and thus our calculations, can terminate with a variety of outcomes. If the accretion rate onto the star is very low, leading to an initial accretion timescale longer than the nuclear burning timescale of the star, stellar evolution is largely unaffected by the AGN disk; we refer to this outcome as ``ordinary.'' If the initial timescale for accretion is shorter than the stellar nuclear burning timescale, the star can accrete substantially before running out of fuel. However, because $L_*\propto M_*^{3-4}$ for $M_*\lesssim10^2 M_\odot$ \citep[e.g.,][]{2013sse..book.....K}, some stars will promptly run out of fuel after reaching high masses; we refer to this outcome as a ``massive'' star. If, even after reaching near-Eddington luminosities, stars can accrete quickly enough to replenish their nuclear fuel, while losing an (on average) equal amount of mass in fusion byproducts via stellar winds, these stars may reach an ``immortal'' state (that is to say, as long as their supply of fuel remains ample). However, if stars accrete more quickly than their structure can adjust, the stellar luminosity and resulting winds will fail to keep pace with accretion, leading to a runaway; gap opening will severely limit the accretion rate and naturally prevent runaway accretion \citep[e.g.,][]{2004ApJ...608..108G}. See Section 2.1 of \citet{2023ApJ...946...56D} for a more quantitative illustration of the preceding reasoning.  

Concerning runaway accretion, it is worth noting that the Kelvin-Helmhotz timescale, $t_{\rm KH}\propto M_*^2(R_*L_*)^{-1}$, or $\propto M_*/R_*$ for massive stars, is a sublinearly increasing function of mass. For Bondi accretion, $M_*/\dot{M}_0\propto M_*^{-1}$; for tidally limited accretion, $M_*/\dot{M}_0\propto M_*^{1/3}$; and after gap opening $M_*/\dot{M}_0\propto M_*^{7/3}$. In the radiative enthalpy-limited regime, the accretion rate becomes roughly independent of density (at constant radiation sound speed) and roughly proportional to $M_*$ through $L_{\rm Edd}$; at higher masses potential energy will dominate accretion feedback, so that $M_*/\dot{M}\propto M_*$. 
Since runaway accretion occurs when $t_{\rm KH}>M_*/\dot{M}_+$, gap opening and (potential energy driven) accretion feedback can reliably truncate runaways, although a temporary runaway phase may occur early in a star's life before it becomes more luminous and while accretion proceeds unlimited. 
For these reasons, we do not terminate calculations considering gap opening or accretion feedback even if they temporarily satisfy $t_{\rm KH}>M_*/\dot{M}_+$. We expect that the predictions of our models will be incorrect while $t_{\rm KH}>M_*/\dot{M}_+$, but that their late-time behavior should be accurate.

Section \ref{sec:fbex} explores how accretion feedback affects these outcomes, compared to considering only radiation-limited Bondi accretion, illustrating how the accretion rate becomes independent of the ambient density in the accretion-feedback-limited regime. Section \ref{sec:grids} explores how accounting for accretion feedback alters stellar evolutionary outcomes across a broad parameter space of conditions drawn from AGN disk models (see Appendix \ref{app:diskModels}). Section \ref{sec:grids} also investigates how accretion feedback alters predictions for the masses and radii of stars in AGN disks, and the rates at which stars can pollute the disk with fusion byproducts. In each of our models, we have assumed initial stellar mass fractions of $X_*=0.74,\,Y_*=0.24$ and disk mass fractions of $X_0=0.72,\,Y_0=0.26.$ Different disk compositions will affect these results only slightly, with higher helium mass fractions requiring higher accretion rates to sustain immortal stars \citep{2023ApJ...946...56D}.

\subsection{The Role of Feedback}\label{sec:fbex}
To illustrate the potential importance of accretion feedback for stellar evolution in AGN disks, we first conduct a small suite of calculations assuming Bondi accretion from the disk ($\dot{M}_0=\dot{M}_B$) for a variety of ambient densities at constant ambient sound speed ($10^6\,{\rm cm\,s^{-1}}$), with and without accretion feedback. Each model assumed an initial stellar mass of $10\,M_\odot$.

\begin{figure}
\includegraphics[width=\linewidth]{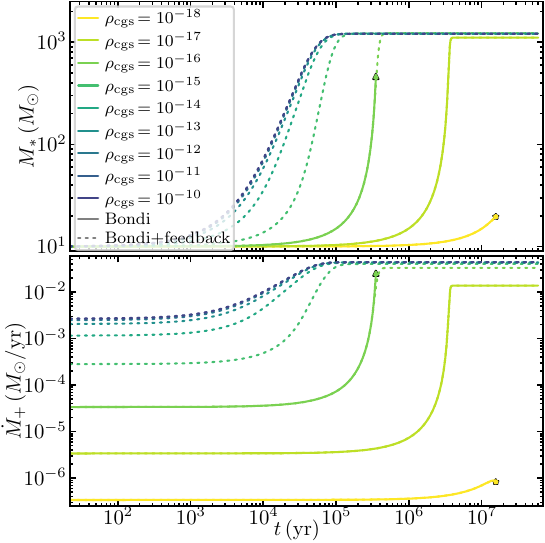}
\caption{Stellar masses and accretion rates assuming Bondi accretion with only radiative feedback (solid lines) and also accounting for accretion feedback (dotted lines) are plotted in the top and bottom panels respectively. Without accretion feedback, most models undergo runaway accretion at these ambient densities.
}
\label{fig:fbExample}
\end{figure}

Figure \ref{fig:fbExample} displays the results of these tests, plotting both the mass and accretion rate for each model. Assuming Bondi accretion with radiative feedback, the lowest-$\rho$ model doubles in mass before running out of fuel, the second-lowest-$\rho$ model achieves an immortal state, and all stars in higher-density media undergo runaway accretion. Considering accretion feedback, the lowest-ambient-density models are unchanged, but the higher-$\rho$ models follow markedly different evolutionary tracks. First, the lower panel of Figure \ref{fig:fbExample} demonstrates that the accretion rate onto the stars becomes independent of density, though still dependent on stellar mass, in the regime probed here. We note that in each of the high-density cases explored here, the star would initially undergo a runaway process, which would eventually terminate as the star grew in mass as described above, before reaching the steady state at which mass loss balances accretion found in these calculations.

\subsection{Feedback in Context}\label{sec:grids}
Accretion feedback can clearly play an important role in shaping the evolution of stars embedded in AGN disks, at least in regions where stellar accretion rates are high. To contextualize the importance of accretion feedback on the evolution of such stars, we have carried out a large suite of calculations, considering Bondi accretion, tidally constrained accretion, and gap-constrained accretion. The parameters for each simulation were drawn from a suite of accretion disk models \citep[][see Appendix \ref{app:diskModels}]{2003MNRAS.341..501S}, which are simply meant to illustrate a plausible range of disk properties. For each set of accretion assumptions, we sampled 167600 parameter combinations, drawn from 200 disk models (varying SMBH mass) and sampled each at 838 radial locations (from 7 to $10^6$ $R_g$, where $R_g\equiv GM_\bullet/c^2$). Each stellar model was initialized at $1\,M_\odot$, and each model was integrated for $10^7$ years, a timescale comparable to the total duration over which an average active galactic nucleus accretes over its lifetime \citep[e.g.,][]{1982MNRAS.200..115S,2002MNRAS.335..965Y}, although individual accretion episodes may be shorter \citep[e.g.,][]{2015MNRAS.451.2517S,2015MNRAS.453L..46K}.

\begin{figure}
\includegraphics[width=\linewidth]{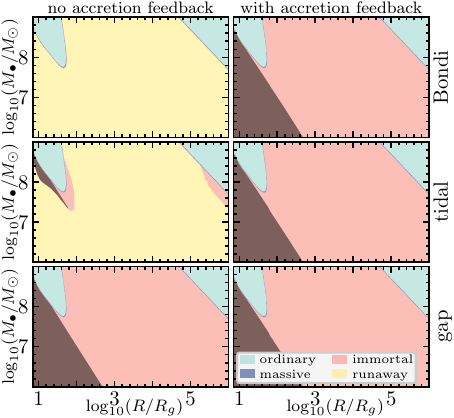}
\caption{The outcomes of stellar evolutionary models across a grid of AGN disk model parameters; the left column plots the results of calculations without accretion feedback, and the right column plots results with accretion feedback; the rows plots results assuming accretion at the Bondi rate, tidally limited accretion, and accretion also limited by gap opening, respectively. The dark gray shaded region in the bottom left corner of each plot indicates models which would be tidally disrupted or stripped of mass by the SMBH based on their radii and the size of their Roche lobe; we do not mark this region for stars that would otherwise undergo runaway accretion since we cannot reliably calculate their masses and radii.}
\label{fig:outcomes}
\end{figure}

To gain a qualitative understanding of the importance of feedback on stellar evolution, we first illustrate the qualitative outcome of each simulation, as described in the introduction to this section, in Figure \ref{fig:outcomes}. For models assuming tidally-limited or Bondi accretion, we categorize any model that satisfies $t_{\rm KH}>M_*/\dot{M}_+$ as undergoing ``runaway'' accretion (see beginning of this section). We classify the remaining models based on their final mass, which is a reasonable proxy for the other evolutionary outcomes: stellar models that reach $10^7$ years with final masses above $100\,M_\odot$ are almost certainly ``immortal'', since the nuclear burning timescale for stars of such masses is $\lesssim10^6$ years; models with final masses less than $8\,M_\odot$ were ``ordinary,'' and the remaining models were ``massive.'' Since at low stellar masses accretion rates typically increase sharply with stellar mass, and because low-mass stars have much longer main sequence lifetimes than the duration of these simulations, whether or not a model exhausted its supply of hydrogen was not useful for judging the stellar evolutionary outcome. 

As expected based on Section \ref{sec:fbex}, accretion feedback prevents models following only Bondi or tidally-limited accretion from undergoing runaways, creating a much wider parameter space for immortal stars. Figure \ref{fig:outcomes} also illustrates how tidally-limited accretion alone can prevent some runaways, but not to nearly the same extent as gap opening and accretion feedback. Notably, when considering gap opening processes, accretion feedback does not make much \textit{qualitative} difference in the outcome of any given calculation, since both processes can terminate runaway accretion. While gap opening should be fairly generic, it does depend somewhat sensitively on assumptions about the stresses and strains within the accretion disk; our parametrization uses an ad hoc \citet{1973A&A....24..337S} viscosity model, and since gap opening depends so strongly on the disk scale height, disks thicker than we have assumed here might make it irrelevant \citep[e.g.,][]{2007MNRAS.375.1070B,2024OJAp....7E..20H}, potentially making accounting for accretion feedback even more important. 

\begin{figure}
\includegraphics[width=\linewidth]{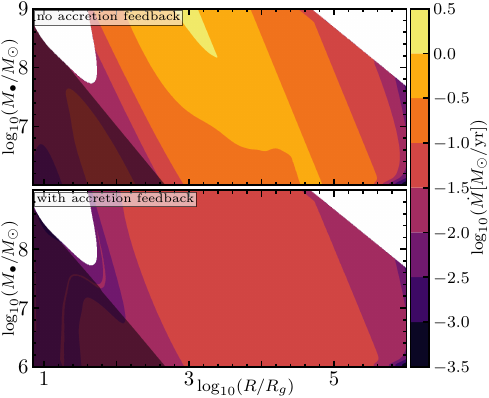}
\caption{Stellar mass-loss rates (equal to accretion rates for these immortal models) over a range of AGN disk model parameters, with and without accretion feedback. Accretion feedback negligibly affects accretion rates onto stellar models subject to lower base accretion rates, but for other models can reduce the accretion rate by more than an order of magnitude.}
\label{fig:gaprates}
\end{figure}

\begin{figure}
\includegraphics[width=\linewidth]{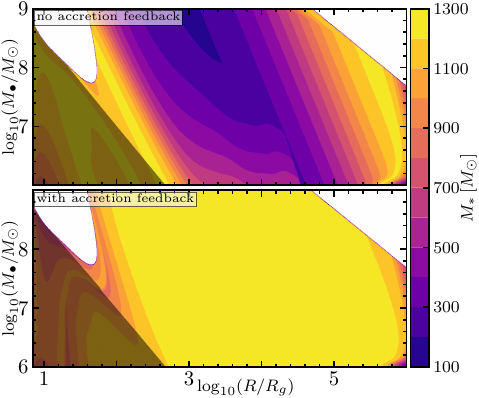}
\caption{The masses of immortal stellar models over a range of AGN disk model conditions, with and without accretion feedback, and under the assumption of gap-limited accretion. Although all of these stellar models were ``immortal,'' neglecting accretion feedback leads to dramatically different stellar properties thanks to the strength of accretion shocks at high accretion rates.}
\label{fig:gapmass}
\end{figure}

\begin{figure}
\includegraphics[width=\linewidth]{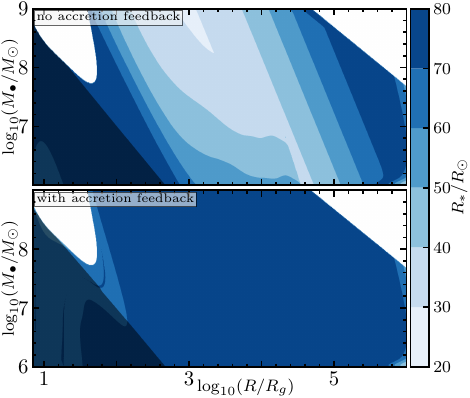}
\caption{The radii of immortal stellar models over a range of AGN disk model conditions, with and without accretion feedback, and under the assumption of gap-limited accretion. Since lower-mass stars have smaller radii, models accounting for accretion feedback tend to predict larger stellar radii.}
\label{fig:gaprads}
\end{figure}

\begin{figure}
\includegraphics[width=\linewidth]{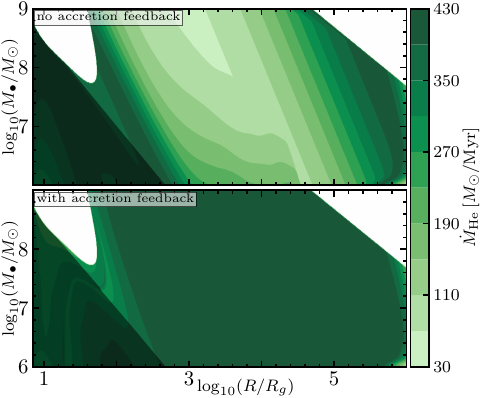}
\caption{The rate at which immortal stellar models pollute their environments with helium over a range of AGN disk model conditions, with and without accretion feedback, and under the assumption of gap-limited accretion. Since lower-mass stars burn hydrogen to helium at lower rates, models accounting for accretion feedback tend to predict higher rates of chemical enrichment.}
\label{fig:gaphe}
\end{figure}

Next we turn to our suites of calculations that accounted for gap opening in the disk, with and without considering accretion feedback. In particular, since ``immortal'' stellar evolution appears to be by far the most common evolutionary outcome under these assumptions, we investigate the properties of these immortal stellar models. 
In Figure \ref{fig:gaprates} we plot stellar mass-loss rates, in Figure \ref{fig:gapmass} we plot their final masses, in Figure \ref{fig:gaprads} we plot their radii, and in Figure \ref{fig:gaphe} we plot the rate at which each star would enrich the disk with helium. 

The direct effect of accretion feedback is visible in the mass-loss rates, and equivalently accretion rates, onto immortal stellar models in Figure \ref{fig:gaprates}. Without including accretion feedback, accretion rates can approach, and even exceed, $1 M_\odot/\rm{yr}$ for a wide range of disk parameters. However, accounting for accretion feedback limits accretion rates to $\lesssim 10^{-1}\,M_\odot/\rm{yr}$. Such high accretion rates are enough to significantly affect stellar structure through the enormous rate at which the accretion shock would deposit energy into the stellar envelope, and thus these differences in accretion rates lead to significant changes in stellar structure. 

As discussed in the final paragraph of Section \ref{sec:basicModel}, and in Appendix \ref{app:escape}, accretion shocks can contribute significantly to the luminosity of the star. When accretion rates are particularly extreme, equilibrium stellar masses can decrease with the accretion rate, and indeed such trends have been present even in the earliest calculations of stellar evolution in AGN disks \citep[e.g.,][]{2021ApJ...910...94C,2021ApJ...916...48D}. Without accretion feedback, accretion rates can become high enough such that the accretion shock can contribute a substantial fraction of the total stellar luminosity, and thus models can reach equilibria between accretion and mass loss at lower stellar masses than they would based on their intrinsic luminosity alone, as seen in the top panel of Figure \ref{fig:gapmass}. 

These lower stellar masses naturally correspond to lower stellar radii, as illustrated in Figure \ref{fig:gaprads}. Accretion feedback prevents stars from reaching these enormous accretion rates, and thus the accretion shock from playing such a significant role. Since accretion feedback is only relevant at very high accretion rates, it has the potentially counterintuitive effect of increasing stellar masses and radii, rather than decreasing them. The higher masses of many immortal stars in the accretion-feedback-limited regime also correspond to higher rates of disk chemical enrichment, as illustrated by Figure \ref{fig:gaphe}, showing the rate of helium production specifically. Our calculations do not track the production of heavier elements directly, but the higher-temperature cores of more massive stars would enrich the disk with heavier elements, particularly carbon, nitrogen, and oxygen, at higher rates. 

\section{Discussion}\label{sec:discussion}

\subsection{Chemical Enrichment}
Estimates of the number of stars present in AGN disks vary considerably. Certain cosmological zoom-in simulations, which find disks predominantly supported by magnetic pressure, suggest that hundreds of stars might form in the outer regions of AGN disks \citep[e.g.,][]{2024OJAp....7E..71H}. On the other hand, models that invoke feedback from stellar populations (and their remnants) to stabilize the outer regions of AGN disks predict hundreds to thousands of stars within a given AGN disk \citep[e.g.,][]{2020MNRAS.493.3732D,2022ApJ...928..191G,2025MNRAS.537.3396E}. Many stars may also be captured into the disk from a surrounding nuclear star cluster \citep[e.g.][]{1993ApJ...409..592A,2020MNRAS.499.2608F}. These stars will gradually enrich the disk with fusion byproducts, particularly byproducts of the CNO cycle. For massive stars, $L_*\propto M_*$, and thus the higher-mass stars allowed by accretion feedback will pollute the disk at higher rates. Since hydrogen-burning stars produce helium at a rate of 
\begin{equation}
\dot{M}_{He}=\frac{L}{27\rm{MeV}/4m_p}\approx 10^{-4}\left(\frac{M_*}{300\,M_\odot}\right)M_\odot/\rm{yr},
\end{equation}
taking $300M_\odot$ is a characteristic mass for immortal stars. Since accretion feedback leads to more massive and luminous stars, enrichment rates of $\sim4\times10^{-4}M_\odot/\rm{yr}$ per star might be feasible. Neglect of accretion feedback could therefore lead to a modest underestimation of disk enrichment rates. 

\subsection{Mergers and Transients}

As we have seen, accretion feedback can allow larger and more massive stars within AGN disks. These properties mean that AGN stars may interact with other stellar-mass objects and the central SMBH in more dramatic ways that would be assumed without accretion feedback: higher stellar radii increase the efficacy of tidal interactions and collisional cross sections, while higher masses provide a larger supply of fuel to passing compact objects.

Extreme AGN flares, the most luminous having energies of $\sim10^{54} \rm{erg}$, are likely caused by tidal disruption events or core-collapse supernovae interacting with a dense surrounding medium \citep[e.g.,][]{2017MNRAS.470.4112G,2021ApJ...920...56F,2026NatAs..10..154G}. Depending on the orbital configurations of disrupted stars, their debris may significantly perturb the inner regions of the disk, especially if higher-mass stars are disrupted \citep[e.g.,][]{2019ApJ...881..113C,2022MNRAS.514.4102M,2024MNRAS.527.8103R}. Larger stars will overflow their Roche lobes at larger distances to the central SMBH, leading to longer timescales for various transient and quasi-periodic phenomena associated with Roche-lobe-overflowing stars and other disk-embedded objects \citep[e.g.,][]{2017ApJ...844...75M,2022ApJ...926..101M}. 

\subsection{Caveats and Future Directions}
As discussed in \citet{2025ApJ...979..245D}, our models assume chemical homogeneity, which makes stellar ``immortality'' much more easily achievable, since fresh accreted material is assumed to instantly mix with spent fuel in the stellar core. This is reasonable for most stars in AGN disks, which should grow to be massive enough to become nearly fully convective, and should spin rapidly \citep{2021ApJ...914..105J}, such that rotation-induced mixing will lead to chemical homogeneity \citep[e.g.,][]{1987A&A...178..159M,2005A&A...443..643Y}. 
However, for stars that accrete very little or accrete material with relatively low angular momentum and accordingly spin up slowly \citep{2021ApJ...914..105J}, this assumption may break down. We plan to test the reasonability of the assumption of chemical homogeneity with future stellar structure and evolution calculations.

We also neglected potentially important time dependencies in other aspects of these calculations. For example, while employing gap-limited accretion we have assumed the instantaneous response of the accretion disk to the gravitational perturbations of the star. 
However, gap opening should require at least a few stellar orbital periods through the AGN disk, which could be considerably longer than stellar evolutionary timescales. 
Our models are also inapplicable to evolution on timescales shorter than the Kelvin-Helmholtz timescale on which a stars luminosity can adjust to changes in its mass. In the preceding calculations when this might occur, we have focused on long-timescale properties of stars in quasi-steady-states. However, such conditions require hydrodynamical modeling, and should be the object of future study.\footnote{Although we calculated time-dependent evolutionary calculations, solving an initial value problem, we remark that the semi-analytical models used here would also be useful to directly search for quasi-steady sates.}

\section{Conclusions}\label{sec:conclusion}
In this work, we have incorporated the effects of radiative enthalpy and potential energy feedback \citep{2024ApJ...974..106C,2025ApJ...987..188C} into a semi-analytical model of stellar structure and evolution \citep{2025ApJ...979..245D}, exploring their impact on stars embedded in AGN disks across a wide range of black hole masses and disk radii. Stars in AGN disks still grow to large masses via accretion; however, we find that the inclusion of accretion feedback qualitatively changes their evolution by suppressing runaway growth, while quantitatively shifting equilibrium stellar masses and radii by over an order of magnitude in some disk regions.

Specifically, feedback tends to limit accretion rates to below $\sim 10^{-1}\,M_\odot\,\mathrm{yr}^{-1}$, rendering the accretion rate nearly independent of ambient gas density and broadening the parameter space over which stars can reach a long-lived ``immortal'' equilibrium \citep{2021ApJ...910...94C,2021ApJ...916...48D}.
These feedback processes also drive significant changes in stellar structure: by limiting the contribution of the shock luminosity $L_S$ to the total energy budget, feedback allows stars to reach higher equilibrium masses $M_*$ while remaining sub-Eddington.
These higher stellar masses correspond to higher intrinsic luminosities from nuclear fusion, such that neglecting accretion feedback could lead one to \textit{underestimate} the rate of chemical enrichment of AGN disks (particularly in helium and CNO-cycle byproducts).

While our semi-analytical approach captures the essential physics of these regimes, several uncertainties remain. The assumption of chemical homogeneity is reasonable for these nearly fully convective and likely rapidly rotating stars, but the time-dependent response of the disk during gap opening and the details of mass loss in the near-Eddington regime require further study. 

Future work utilizing full 1D stellar evolution calculations \citep[e.g., using \texttt{MESA};][]{2011ApJS..192....3P} will be essential for refining these evolutionary tracks and explore the interplay between feedback and internal stellar structure in greater detail.

\section*{Software}
\texttt{matplotlib} \citep{4160265}, \texttt{numpy} \citep{5725236}, scipy \citep{2020SciPy-NMeth}

\section*{Acknowledgments}

AJD is grateful for the hospitality of the Center for Computational Astrophysics on numerous occasions, and to Yan-Fei Jiang and Mor Rozner for stimulating discussions.
The Center for Computational Astrophysics at the Flatiron Institute is supported by the Simons Foundation.
Support for this work was provided by NASA
through the NASA Hubble Fellowship grant No. HST-HF2-
51553.001, awarded by the Space Telescope Science Institute, which is operated by the Association of Universities for Research in Astronomy, Inc., for NASA, under contract NAS5-26555. 

\bibliographystyle{aasjournal}
\bibliography{references}

\appendix
\section{Semi-Analytical Model Updates: Escape Velocity}\label{app:escape}
At very high accretion rates, the rate of energy exchange between the accretion stream and stellar surface can approach, or even exceed, the intrinsic luminosity of the star. Taking a naive estimate of the accretion shock luminosity, 
\begin{align}
L_S&=\dot{M}_+GM_*/R_*  \label{eq:ls1} \\ 
&\approx3.1\times10^7\left(\frac{M_*}{M_\odot}\right)\left(\frac{R_\odot}{R_*}\right)\left(\frac{\dot{M}_+}{M_\odot/\rm{yr}}\right)\!L_\odot\\
&\approx \left(\frac{2}{1+X}\right)\left(\frac{R_\odot}{R_*}\right)\left(\frac{\dot{M}_+}{M_\odot/\rm{yr}}\right)10^4L_{\rm Edd}.
\end{align}
Thus, for the masses, radii, and accretion rates for stars in AGN disks (Figures \ref{fig:gaprates}--\ref{fig:gaprads}), the accretion shock might dramatically alter the structure of the star. However, with such strong radiation fields, the notion of escape velocity used to construct the above estimates breaks down. To correct this deficiency, which is imperceptible at lower accretion rates, we instead estimate
\begin{equation}
v_{\rm esc}^2 = \frac{2GM_*}{R_*}\left(1-\frac{L_*}{L_{\rm Edd}} - \frac{L_S}{L_{\rm Edd}}\right),
\end{equation}
in which case the accretion shock luminosity becomes
\begin{equation}\label{eq:ls2}
L_S = \dot{M}_+\frac{GM_*}{R_*}\!\left(1\!-\!\frac{L_*}{L_{\rm Edd}}\right)\left(
1\!+\!\frac{\dot{M}_+}{L_{\rm Edd}}\frac{GM_*}{R_*}\right)^{-1}\!\!\!\!\!.
\end{equation}

\begin{figure}
\includegraphics[width=\linewidth]{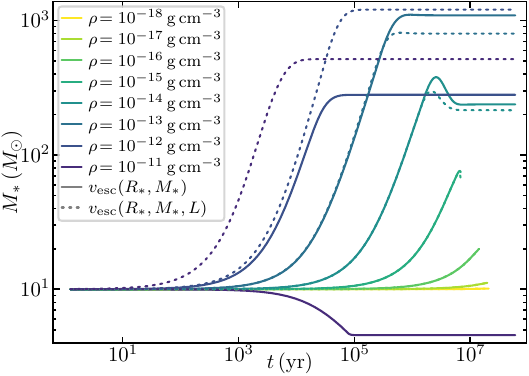}
\caption{The masses over time of stellar model over a range of densities, comparing the use of Equation (\ref{eq:ls1}, solid lines) and Equation (\ref{eq:ls2}, dotted lines). These calculations neglected radiation feedback and assumed tidally-limited accretion, fixing $\Omega=10^{-8.5}\,\rm{s}^{-1}$ and $c_s=10^6\,\rm{cm}\,\rm{s}^{-1}$ for all calculations. 
}
\label{fig:vesc}
\end{figure}

We illustrate the behavior of models assuming Equation (\ref{eq:ls1}) versus Equation (\ref{eq:ls2}) in Figure \ref{fig:vesc}, assuming tidally-limited accretion and neglecting accretion feedback for simplicity. At lower densities (and thus accretion rates) the two assumptions produce virtually identical results; at a density of $\rho\sim10^{-15}\,\rm{g\,cm^{-3}}$ the mass reached by the immortal stellar models differs visibly, with differences becoming more significant at higher accretion rates; at $\rho\sim10^{-11}\,\rm{g\,cm^{-3}}$ the evolution of the model assuming Equation (\ref{eq:ls1}) becomes qualitatively erroneous, the accretion luminosity driving outflows strong enough to overwhelm accretion entirely. Besides the exemplary calculations in Figure \ref{fig:vesc}, we use only Equation (\ref{eq:ls2}) throughout this paper. 

\section{Disk Models}\label{app:diskModels}
In the following, we briefly describe the AGN disk models used in this work. In short, we use the $\alpha-$disk form of the \citet{2003MNRAS.341..501S} equations (that is to say, setting the viscosity to $\nu=\alpha c_s^2/\Omega$), along with the tabulated opacities constructed by \citet{2021MNRAS.508..453Z}.\footnote{That work combined atomic opacities from \citet{2016ApJ...817..116C}, molecular opacities from \citet{2014ApJS..214...25F}, and dust opacities from \citet{2018ApJ...869L..45B} to produce an opacity table with coverage over a wide range of temperatures and densities; their opacity tables are available at \url{https://github.com/zhuzh1983/combined-opacity}.}
The equations of the model are given by
\begin{align}
T^4&=T_{\rm eff}^4(\frac{3}{8}\tau + \frac{1}{2} + \frac{1}{4\tau}) \label{eq:SG1}\\
\tau &= \rho H \kappa \label{eq:SG2}\\
h\rho c_s^2 &= \frac{\dot{M}'\Omega}{3\pi\alpha} \label{eq:SG3}\\
\rho c_s^2 &= \frac{\rho k T}{\mu} + \frac{\tau \sigma}{2c}T_{\rm eff}^4 \label{eq:SG4}\\
c_s &= H\Omega, \label{eq:SG5}
\end{align}
where $\kappa(\rho,T)$ is given by an interpolation over the tables provided by by \citet{2021MNRAS.508..453Z} (specifically a bicubic spline of $\log{\kappa}$ as a function of $\log{\rho}$ and $\log{T}$). The preceding equations are closed by either
\begin{equation}
\sigma T_{\rm eff}^4=\frac{3}{8\pi}\dot{M}'\Omega^2
\end{equation}
when the disk is gravitationally stable ($\Omega^2>2\pi\rho G$), or by 
\begin{equation}
\rho = \frac{\Omega^2}{2\pi G}
\end{equation}
otherwise. 

We take $\Omega$ to be the Keplerian value for each central black hole mass, $\Omega^2=GM_\bullet/r^3$, and set $\dot{M}'=\dot{M}(1-\sqrt{r_0/r})$, where $r_0\equiv 6GM_\bullet/c^2$, which sets the angular momentum current through the disk based on the specific angular momentum near the innermost circular orbit around the central black hole. For each model we set the accretion rate to $\dot{M}=4\pi G M_\bullet m_H/(c\sigma_T)$ (where $\sigma_T$ is the Thompson electron scattering cross section), which corresponds the central black hole accreting at $10\%$ of the Eddington-limited rate while converting $10\%$ of the accreted rest mass-energy to radiation. We also set the viscosity coefficient to $\alpha=0.1$. 

\begin{figure}
\includegraphics[width=\linewidth]{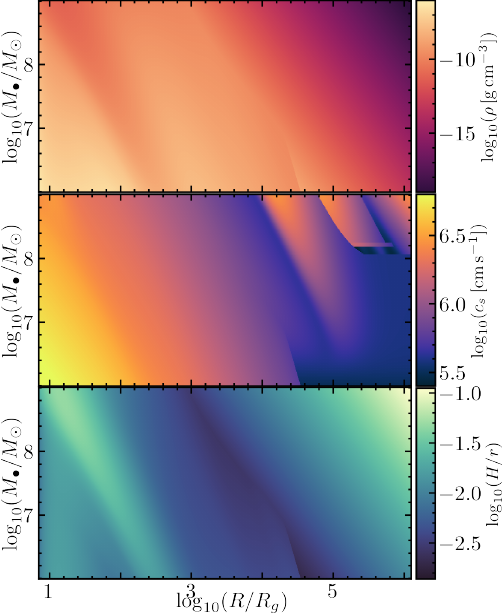}
\caption{The disk models used in this work, but more specifically quantities most directly related to stellar evolution: the top panel plots the gas density, the middle panel plots the total sound speed, and the bottom panel plots their aspect ratio ($H/R$), all as functions of central SMBH mass $M_\bullet$ and radial location. At higher masses the disk transitions from the $Q>1$ equations to the $Q=1$ equations (where $\rho$ is a simple power law $\propto\Omega^2$) at about $10^3 R_g$, while at lower masses the transition happens at closer to $10^{4.5}\,R_g$. The numerous sharp features in the sound speed, particularly in the outer disk, result from sharp changes in the opacity such as dust sublimation.}
\label{fig:dmods}
\end{figure}

For the purposes of this investigation, we calculate models for central black hole masses $M_\bullet/M_\odot\in[10^6,10^9]$ and the radial range $r/r_g\in[10, 10^6]$, where $r_g\equiv GM_\bullet/c^2$. The resulting density, sound speed, aspect ratio ($h/r$) profiles are shown in Figure \ref{fig:dmods}. These are the main inputs to the calculations in Section \ref{sec:results}. 

\end{document}